\documentclass[12pt]{iopart}

\usepackage{float}
\usepackage{epsfig}
\usepackage{cite}
\usepackage{caption}
\usepackage[usenames, dvipsnames]{color}
\usepackage{iopams}  
\begin{document}

\title{Consequences of nonreciprocity in reflection from a truncated spatial Kramers-Kronig medium}

\author{Sudipta Saha$^1$, K. V. Sowmya Sai$^2$, Nirmalya Ghosh$^{1,*}$ and Subhasish Dutta Gupta$^{2,**}$}

\address{$^1$Department of Physical Sciences, Indian Institute of Science Education and Research Kolkata, Mohanpur-741 246, India}
\address{$^2$School of Physics, University of Hyderabad, Hyderabad 500046, India}
\ead{$^*$nghosh@iiserkol.ac.in; $^{**}$sdghyderabad@gmail.com}
\vspace{10pt}

\begin{abstract}
	We study the truncated Kramers-Kronig (KK) medium to reveal the effects of nonreciprocity in reflection on the scattering and the wave guiding properties. We focus on the simplest possible spatial profile mimicking the dielectric response. We show that a finite slab of KK medium can support only one sided null scattering under bidirectional identical illumination. However, for a phase lag between the two incident beams it is possible to have null scattering on the other side. This is shown to be a consequence of the nonreciprocity in reflection due to the inherent asymmetry of the KK medium. We further look at the possibility of surface modes for unidirectional illumination.  For specific range of parameters, we show the existence of TM polarized surface plasmons only for illumination from one side. The origin of such modes is traced to the inhomogeneous metal like layer with negative real part of the permittivity in the KK medium.
%
%
\end{abstract}

%
%
%

%
%

\section{Introduction:}
Suppression of reflection has been one of the central topics in optics for several decades mostly because of its potential for various applications ~\cite{bornwolf,Leknerbook,poitras2004}. Inhomogeneous dielectric stratified media (both continuous and discrete) have been explored to this end \cite{Kiriuscheva,lekner2007reflectionless,Baryshnikova2015}. Complex dielectric profiles have also been studied \cite{Horsleywave2016,HorsleyKdV2016}. In the pioneering work by Kay and Moses\cite{kay1956} on reflectionless potentials and its later theoretical and experimental follow up \cite{lekner2007reflectionless,sdgoe2007,sdgoe2014}, the dielectric index profile was assumed to be lossless. Recently, Horsley \etal \cite{Horsleynatph2015} have introduced a general class of spatially inhomogeneous planar	dielectric medium, namely, the spatial Kramers-Kronig medium, which offers unidirectional anti-reflection\cite{LonghiOL2015,LonghiEPL2015,Mostafa2014}. For this new class of locally isotropic, non-magnetic medium, the spatial profiles of the real and the imaginary parts of the complex dielectric permittivity $\epsilon(z)$ are related by the Kramers-Kronig (KK) relations\cite{Horsleynatph2015}. The spatial KK relations will hold for any planar dielectric medium whose spatial permittivity profile $\epsilon(z)$ is an analytic function (that is without poles) of complex spatial co-ordinate ($z=z^\prime+iz^{\prime\prime}$) in the upper (or lower) half of the complex plane \cite{Horsleynatph2015}. As can be anticipated from the response theory\cite{waveopticsbook}, for such media, the reflection of wave from the left (or the right) of the profile would vanish for any given angle of incidence\cite{Horsleynatph2015,LonghiEPL2015}, thus offering the possibility of unidirectional antireflection. Unidirectional invisibility has also been observed in PT symmetric and other systems \cite{Longhi2011,Feng2013,prlLin2011}, some of which fall under the category of the KK media\cite{Horsleynatph2015}. A deeper look at the underlying physics and mathematics of the KK medium has revealed the possibility of bidirectional invisibility for incidence from either the left or right\cite{Longhi_bidirection}. The novelty of the spatial KK media as compared to other non-reflecting profiles is that they do not require any symmetry under spatial or temporal inversion. Moreover, they also do not require presence of gain or negatively refracting media to produce anti-reflection behavior or invisibility. Despite the considerable promise of this new class of optical media, experimental realization of the KK dielectric function profile still poses a formidable task, since one needs to simultaneously	engineer the spatial distribution of both the real and imaginary parts of the dielectric function. Moreover, like in the case of reflectionless potentials, most of the above results hold for KK media of infinite extent \cite{Horsleynatph2015}. Thus, one needs to consider the effect of truncation of the profile, which is inevitable for any practical realization. In the context of reflectionless potential it was shown that truncation of the profile leads to loss of antireflection for grazing and near-grazing incidence \cite{sdgoe2007}. Some of the consequences of truncation was addressed in the original work of Horsley \etal\cite{Horsleynatph2015} (see the supplementary material). In this paper we study the truncated profile with two specific targets. Firstly, we probe the possibility of coherent perfect absorption (CPA) in a finite KK medium for simultaneous bidirectional illumination. Note that CPA in linear \cite{wan2011,sdgOE2012,sdgOL2012,waveopticsbook,sdgOL2007} and nonlinear \cite{nireekshan_absorption2013,reddyCC2013,gapsoliton2014} systems as well as its quantum variations \cite{twophotonSDG,gsa_cpa,Roger2015} have been studied in detail in recent years. We show that the inherent asymmetry in the KK medium rules out the possibility of having CPA though one can have complete suppression of scattering only on one side. This is shown to be a direct consequence of `nonreciprocity in reflection' in an asymmetric structure with loss or gain \cite{receprocity_sdg2002,sdg_joptb2004,waveopticsbook}. The second part of our paper focuses on the ability of a symmetrically loaded KK medium to support plasmonic surface waves. To this end we exploit a highly truncated KK medium where there is a metal like (with negative real part of the dielectric function) section. The surface mode is shown to exist only for TM polarization and when the metallic part faces the incident beam. The mode can easily be detected in the attenuated total reflection (ATR) geometry \cite{waveopticsbook}. Such modes do not show up when the illumination is from the opposite dielectric side. Despite having an inhomogeneous distribution of the dielectric function the mode is shown to be localized at the surface with expected local field enhancement.
\section{One sided complete suppression of scattering }
\begin{figure}[t]
\centering	
\includegraphics[width=0.65\linewidth]{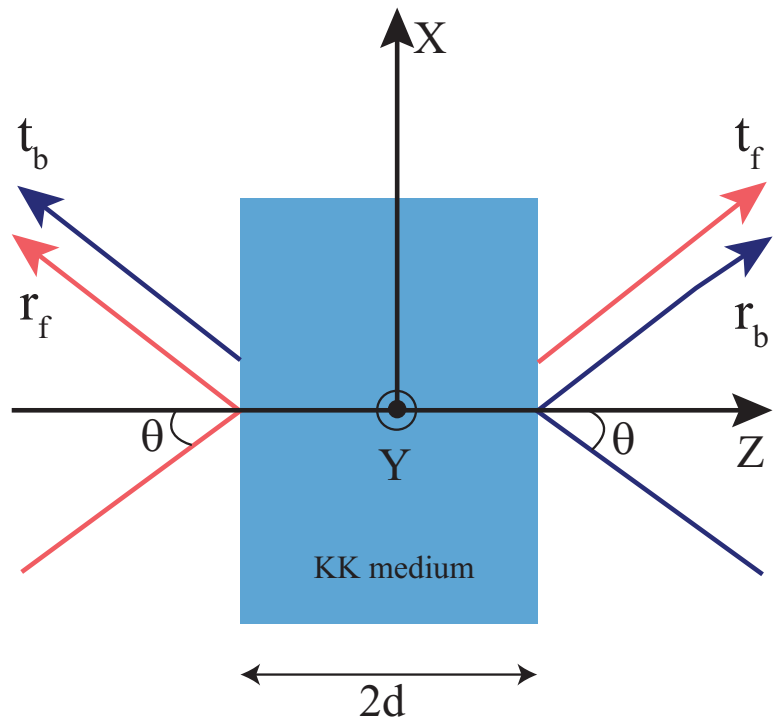}
\caption{\label{fig1} (Color online) Schematic view of the finite KK medium embedded in a dielectric medium with dielectric constant $\epsilon_i$.}
\end{figure}
Consider the structure shown in Fig. \ref{fig1} where a KK medium of finite extent 2d is embedded in a dielectric medium with dielectric constant $\epsilon_{i}$. We pick the simplest possible KK profile given by
\begin{equation}
\epsilon(z) = \epsilon_{b}-\frac{A}{z/\xi+i}
\end{equation}
where A and $\xi$ are real constants and $\epsilon_{b}$ gives the background dielectric constant. The origin is chosen at the centre of the KK medium occupying the domain -d$\leq$ z $\leq$ d. All the media are assumed to be nonmagnetic. In contrast to the existing studies with unidirectional illumination we let the medium be excited by two monochromatic TE-polarized plane waves from the opposite sides at an angle $\theta$. Left to right(right to left) incident and resulting reflected and transmitted amplitudes are labeled by subscript `f'(`b'). The illumination geometry is the one used for CPA \cite{sdgOE2012}. By virtue of the spatial symmetry of the standard CPA systems with identical illumination $r_{f} = r_{b}$ and $t_{f} =t_{b}$ and the CPA condition reduces to $|r+t|=0$ ($r$ and $t$ represent the amplitude reflection and transmission coefficients). This amounts to the equality of the magnitudes of the amplitude reflection and transmission coeffecients and a mutual phase difference of $\pi$ between them ensuring complete destructive interference between the two on either side. In our case of finite KK medium the inherent asymmetry requires the fulfilment of both the following relations
\begin{eqnarray}
|r_{f}+t_{b}|= 0, \label{eq2}  \\ 
|r_{b}+t_{f}|= 0. \label{eq3}
\end{eqnarray}
\begin{figure}[t]
\includegraphics[width=\linewidth]{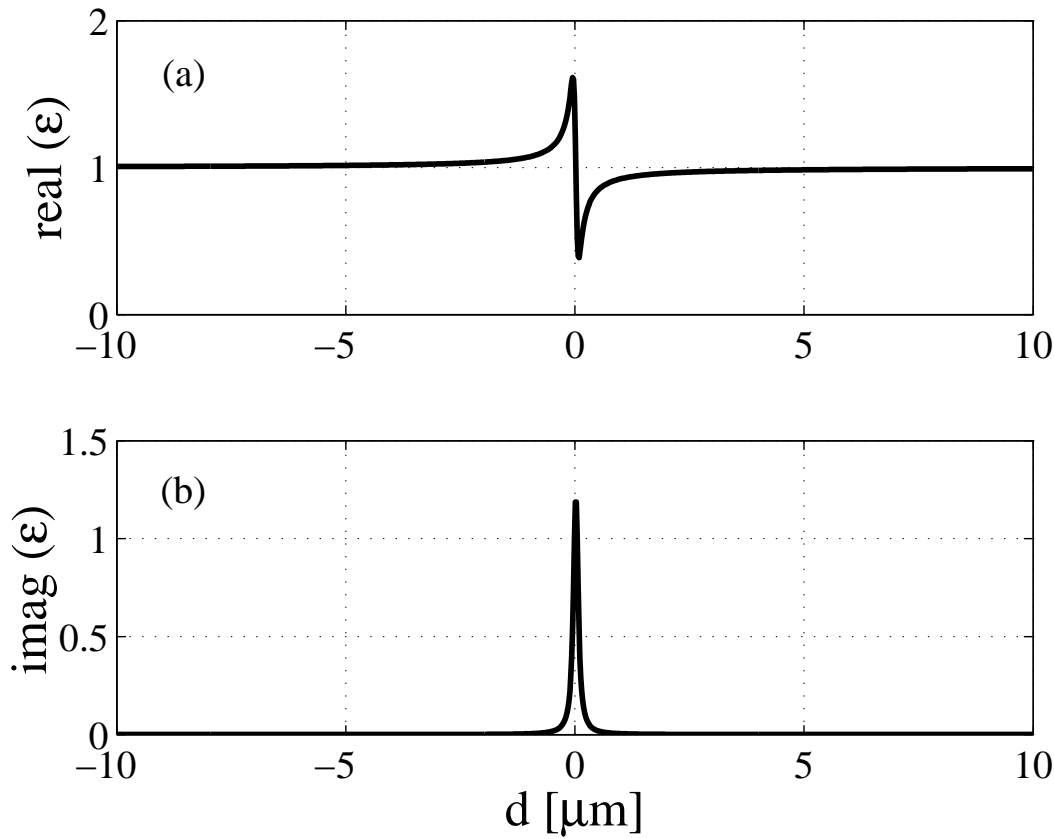}
\caption{\label{fig2} (a)Real and (b) imaginary parts of KK permittvity profile inside a slab of thickness 20 $\mu$m with $A=1.25$, $ \xi=0.06\;\lambda$, $\lambda=1.06\;\mu$m and $\epsilon_b=1$.}
\end{figure}
It is now well understood that irrespective of the spatial symmetry, transmission is always reciprocal (same for both forward and backward illumination) \cite{receprocity_sdg2002}. The same is not true for reflection for lossy (gain) systems for identical incident fields. It has different magnitudes and in general $|r_{f}|\neq|r_{b}|$. Thus it is practically impossible to simultaneously fulfil both the conditions given by Eq. (\ref{eq2}) and Eq. (\ref{eq3}) for any chosen set of system parameters. In other words complete destructive interference on both sides of the KK medium is not possible and a finite KK medium cannot absorb all the incident light. Though CPA is ruled out, one can still have complete suppression of scattering only on one side of the medium by fulfilling one of the above equations for a particular set of parameters.

 \begin{figure}[h]
 \includegraphics[width=\linewidth]{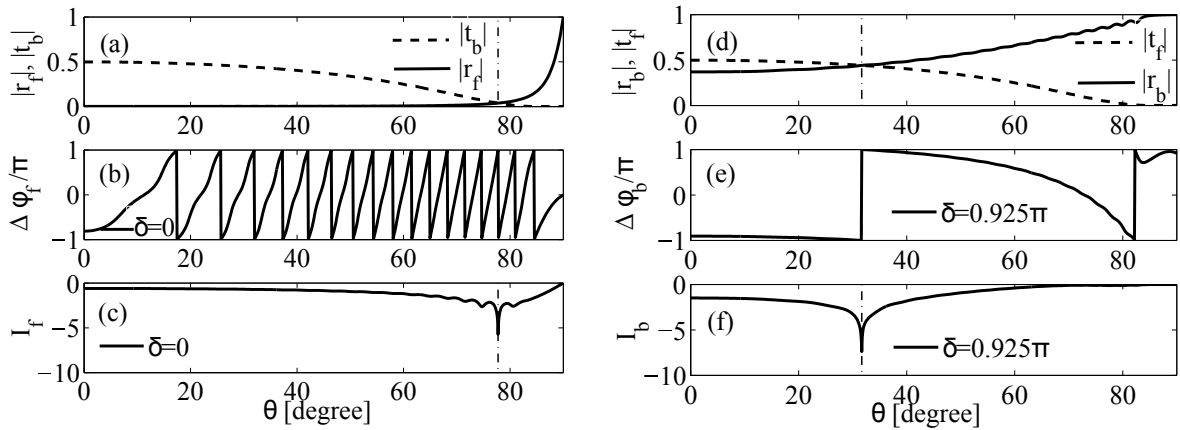}
 \caption{  \label{fig3} (a) Magnitude of the amplitude reflection $|r_f|$ (solid curve) and transmission $|t_b|$ (dashed curve) coefficients on the left side as a function of angle of incidence $\theta$. (b) Angle dependence of the phase difference $\Delta \phi_f$ between $r_f$ and $t_b$ with no phase lag ($\delta=0$) between the incident fields. (c) Angle dependence of total scattering on the left  $I_f=\log_{10}|r_f+t_b|^2$ for $\delta=0$. In Figs. (a),(b) and (c), the vertical dashed-dotted line at $\theta=77.75^\circ$ marks the angles of incidence for null scattering on the left. Analogous quantities, albeit on the right side for $\delta=0.925 \pi$, are shown in Figs. (d), (e) and (f), respectively. The black dashed-dotted line at $\theta=31.68^\circ$ marks the angles of incidence for null scattering on the right. Note that the reciprocity in transmission coefficients ensures $|t_f|=|t_b|=|t|$. The parameters used for numerical calculations are as follows: $d=10\mu$m, $A=1.25$, $ \xi=0.06\;\lambda$, $\lambda=1.06\;\mu$m and $\epsilon_b=1$.}
 \end{figure}
We would like to note here that ‘nonreciprocity’ in a broader sense refers to systems that break the Lorentz reciprocity condition \cite{Jalas_2013,Potton_2004}.  It has been shown that the Lorentz reciprocity can be broken in specific systems (e.g., exhibiting nonlinearity, having asymmetric permittivity or permeability tensor and so forth \cite{Potton_2004,Jalas_2013,PTsymm_sdg}) leading to nonreciprocal transmission, which is desirable for potential applications involving optical diode action and isolation \cite{Jalas_2013,PTsymm_sdg}. In contrast, nonreciprocity in our case (in the context of the one dimensional stratified KK-medium) refers only to the reflected waves \cite{receprocity_sdg2002}. The truncated spatial KK medium studied by us satisfies the Lorentz reciprocity condition, as in this case, the dielectric permittivity of the medium is a scalar. As a consequence of Lorentz reciprocity, transmission is always reciprocal (optical diode action or isolation is ruled out in such structures). Interestingly, as demonstrated previously \cite{receprocity_sdg2002,Potton_2004,PTsymm_sdg,epjap_2016,sdg_joptb2004,Kumari_2012,Woerdman_2003}, for lossy systems lacking spatial symmetry reflection can be nonreciprocal.
\par	
Numerical calculation of the reflection and transmission coefficients is carried out by a fine subdivision of the KK medium and assuming the optical properties (dielectric constant) to be uniform in each sublayer. A characteristic matrix approach \cite{bornwolf,waveopticsbook} is invoked to calculate the transmission and reflection spectra. For the results shown in Fig. \ref{fig3}, the following set of parameters were chosen for calculation: $A=1.25$, $\xi=0.06\;\lambda$, $\epsilon_{b}=1$, $\epsilon_{i}=\epsilon_{f} =1$, $\lambda=1.06\;\mu m$, $d=10\; \mu m$. It is important to note that the relative phase $\delta$ of the incident beams, which can be easily controlled by additional delay on the path of one of the beams, plays a very crucial role in determining the total scattering on either side of the KK medium. We have defined the scattering to the left and right of the KK profile as $I_f=\log_{10}|r_f+t_b|^2$ and $I_b=\log_{10}|r_b+t_f|^2$, respectively. The corresponding phase differences for left and right incidence are denoted by $\Delta\phi_f= \arg(r_f)-\arg(t_b)$ and $\Delta \phi_b=\arg(r_b)-\arg(t_f)$, respectively. The absolute values of the reflection and transmission coefficients, their phase difference $\Delta\phi_f$ and the total scattering on the left are shown in Figs. \ref{fig3}(a), (b) and (c), respectively. It is clear from these figures that for the chosen set of parameters and for $\delta=0$, the magnitude of $r_{f}$ and $t_{b}$ match at $\theta=77.75^\circ$, with $\Delta\phi_f=\pi$ resulting in a dip in total scattering on the left of the structure ($ |r_{f}+t_{b}|=0$). The right panel of Fig. \ref{fig3} shows similar quantities on the right side of the KK profile. For example, Fig. \ref{fig3}(d) shows the reflection and transmission coefficients on the right side of the structure. It is interesting to note that for the said parameters with $\delta=0$, Eq. (\ref{eq3}) can not be satisfied since the phase relation can not be met at $\theta=31.68^\circ$, though $|r_{b}|=|t_{f}|$. However, as shown in Figs. \ref{fig3} (e) and (f), a delay of one of the beams amounting to a phase difference $\delta=0.925\pi$ can lead to null scattering on the right of the structure as well. It is thus clear that a controlled delay of the two incident beams can engineer the nature of scattering on the left and right of the structure.  For a chosen set of parameters, one can have complete suppression of scattering only on one side of the KK profile, though CPA is ruled out.
\par
\begin{figure}[t]
\centering
\includegraphics[width=\linewidth]{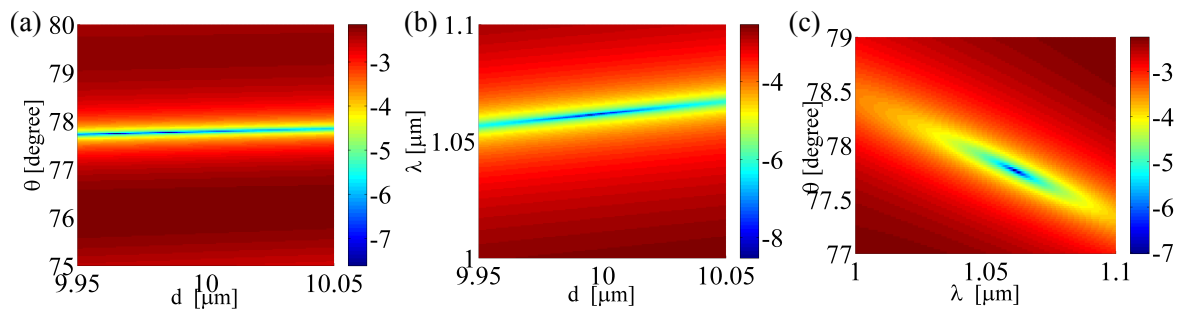}
\caption{\label{fig4} (Color online) Parameter dependence of the scattering ($I_f$) on the left of the KK medium is shown: (a) $I_f$ as a function of $\theta$ and d for $\lambda=1.06\;\mu m$, (b) $I_f$ as a function of $\lambda$ and d for $\theta=77.75^\circ$ and (c)  $I_f$ as a function $\theta$ and $\lambda$ for $d=10\;\mu m$. The parameters used are the same as in Fig. \ref{fig3}.}
\end{figure}

The parameter dependence of the scattering cancellation dips of $|r_{f}+t_{b}|$ is shown in Figs. \ref{fig4}(a), (b), (c) for the same set of parameters used in Fig. \ref{fig3}. It is clear from Figs. \ref{fig4}(a), (b), (c) that there are distinct regions in the parameter space where the distructive cancellation of the scattering can take place. The system parameters control the crossing of the magnitudes of the reflection coefficient (for forward and backward incidence) with that of the transmission coefficient, while the regulated delay can ensure suppression of scattering on the desired side. Note that the crossings are not always possible leading to only one-sided supression of scattering. For eample, for $A=2.0$, there is no matching of $|r_b|$ and $|t_f|$ leading to finite scattering on the right of the structure.
\section{Surface modes of a truncated KK medium}
\begin{figure}[t]
\includegraphics[width=\linewidth]{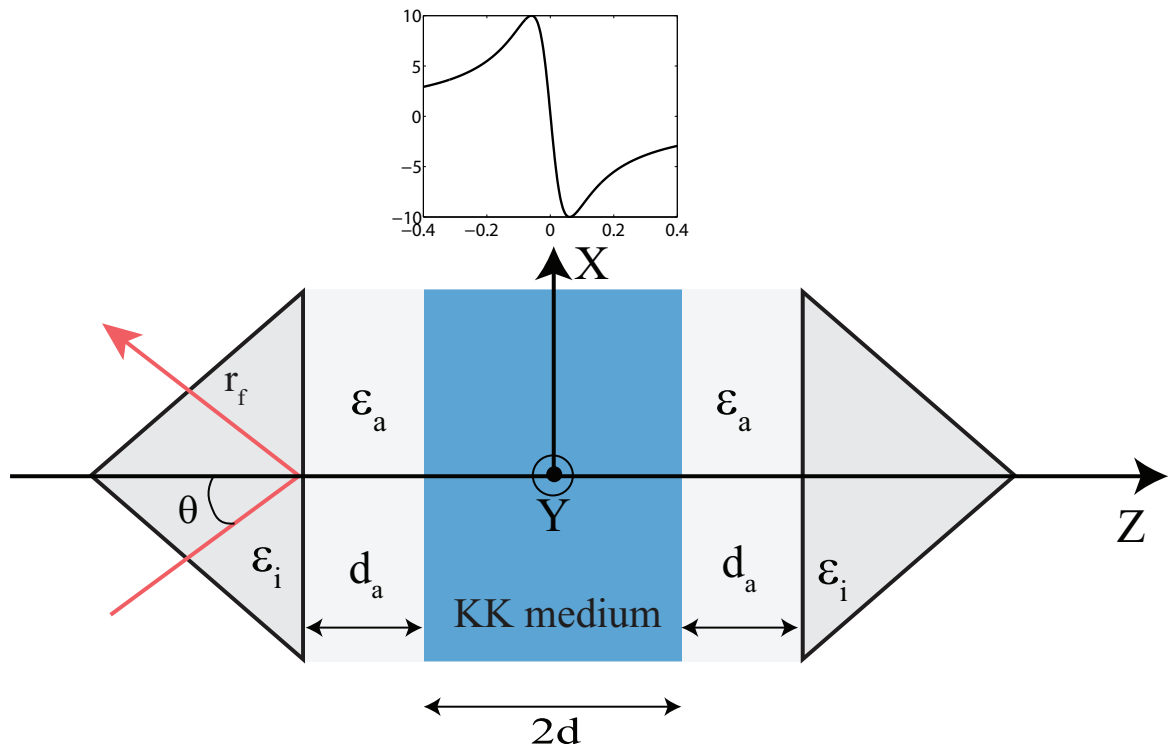}
\caption{\label{fig5} (Color online) Schematic of the folded Kreschtmann configuration (prism-air-KK medium-air-prism) for realization of ATR. The real part of $\epsilon(z)-\epsilon_b$  for the KK medium is shown in the inset. }
\end{figure}
\begin{figure}[t]
	\includegraphics[width=\linewidth]{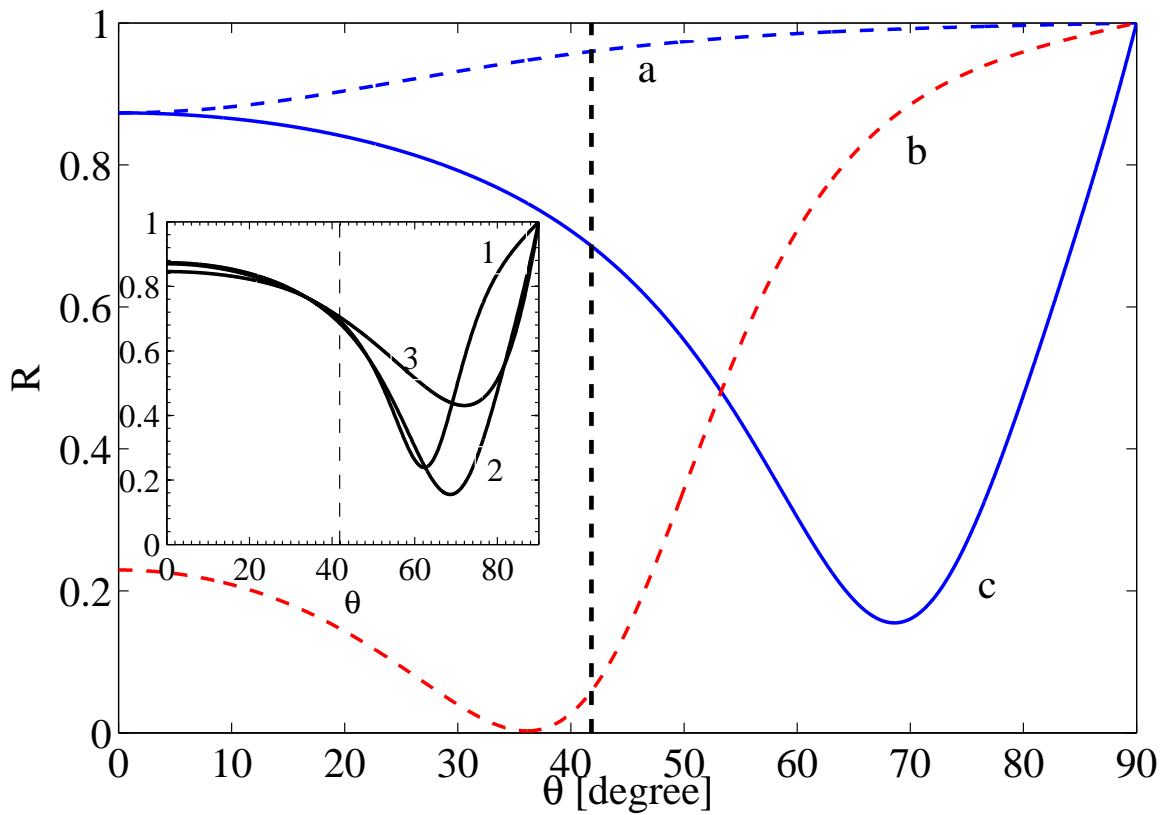}
	\caption{\label{fig6} (Color online) Comparison of intensity reflection coefficient $R$ as a function of the angle of incidence $\theta$: the blue (red) dashed line marked by `a' (`b') is for the incidence of TE (TM) polarized light from the right (left) of the KK medium. The excitation of the surface plasmon is shown by the blue solid line marked by `c' is for the TM polarized light incident from the right side of the KK medium. The inset shows $R$ for varying thickness $d_a$ of the spacer layer for right incident TM wave. `1',`2' and `3' are for $d_a=0.28\mu m,~0.20\mu m ~\rm{and}~0.12\mu m$, respectively. The vertical black dashed line marks the critical angle for high index prism/spacer layer interface. The parameters for the numerical calculations are as follows: $A=20$, $\epsilon_b=1$, $\xi=0.06\;\lambda$, $d=0.4\;\mu m$, $\epsilon_i=\epsilon_f=2.25$, $\epsilon_a=1$ and $\lambda=1.06 \mu m$. }
\end{figure}
\begin{figure}[h]
\includegraphics[width=\linewidth]{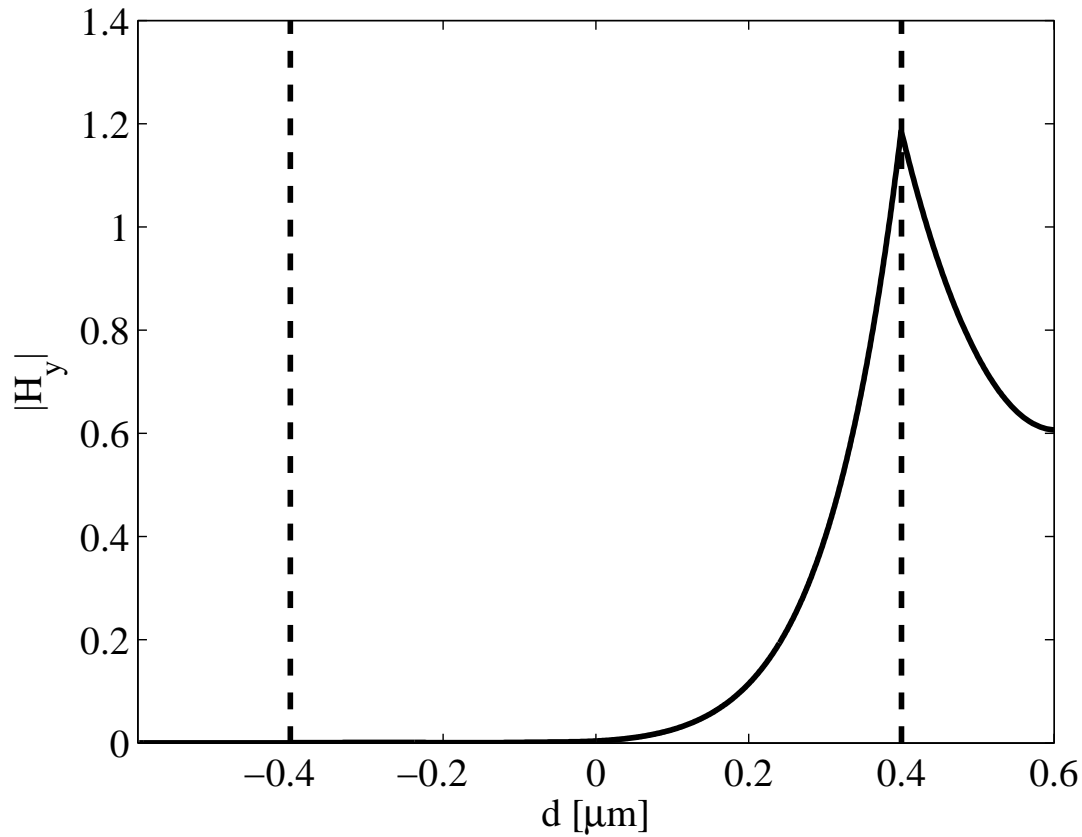}
\caption{\label{fig7} Absolute value of the tangential to the surface component of the magnetic field $|H_y|$ inside the layered (air-KK medium-air) medium.  The vertical dashed lines represent the interfaces between different media. The parameters are the same as in Fig. \ref{fig6}.}
\end{figure}

In this section we probe for the surface modes of a highly truncated KK medium in a folded Kreschtmann configuration (prism-air-KK medium-air-prism) \cite{waveopticsbook} (see Fig. \ref{fig5}). Let the system be illuminated by a plane TM-polarized light from one side. As mentioned earlier, the asymmetry of the KK medium ensures distinct reflection spectra for incidence from left and for incidence from right. In what follows, we show that for left incidence one cannot excite the surface mode, while for right incidence one has a well defined dip in the ATR spectra. For numerical calculations we have used the following set of parameters: $A=20$, $\epsilon_b=1$, $\xi=0.06$, $d=0.4\mu m$, $\epsilon_i=\epsilon_f=2.25$, $\epsilon_a=1$, $\mu_i=\mu_f=1$ and $\lambda=1.06 \mu m$. The results for the intensity reflection coefficient as a function of angle of incidence is shown in Fig. \ref{fig6}, where we have presented the case of TE polarized plane wave also just to highlight the plasmonic origin of the surface mode. The inset shows how the excitation of the mode can be optimized by controlling the width of the spacer layer $d_{a}$. It is clear from Fig. \ref{fig6}, that the surface mode can be excited only for incidence from right. This can be easily explained by looking at the dielectric function profile shown in the inset of Fig. \ref{fig5}. For large $A$, left half of the KK medium acts like a dielectric, while the major portion of the right half behaves like a metal with $Re(\epsilon(z))<0$. 
Thus the right interface between the KK medium and the spacer layer essentially represents a metal-dielectric interface supporting the surface plasmon localized at that interface. The left interface, being that between two dielectrics cannot support the surface mode, which explains unidirectional excitation. In order to verify this, we plotted the field distribution inside the layered medium (see Fig. \ref{fig7}) which clearly shows the localization of the field and the associated local field enhancement. It is interesting to note the exponential decay of the field profiles away from the interface though we have a inhomogeneous `metal' /dielectric interface instead of the standard uniform metal/dielectric interface for surface plasmons.
\section{Conclusions}
We have studied a truncated KK medium to reveal two of its interesting properties. Under simultaneous bidirectional illumination we have ruled out the possibility of having CPA in such structures. However, the complete suppression of scattering on either side of the structure is shown to be feasible for very specific system parameters. We have shown how the phase delay between the incident beams can control the suppression of scattering on either side of the structure. In addition, we probed the feasibility of surface plasmons in a highly truncated system showing that such modes can exist only for incidence from the `metallic' side. Both the aspects discussed above owe their origin to the nonreciprocity in reflection in a generalized lossy system lacking spatial symmetry. \\
We would like to note that materials with dielectric permittivity profile satisfying spatial Kramers-Kronig relation have not yet been realized. This is primarily due to the inherent challenge in simultaneous engineering of the spatial distribution of both the real and the imaginary parts of the dielectric function. Despite this, theoretical studies on such spatial KK medium have attracted lot of recent attention due to its fascinating optical properties and its  promise to offer new ways for the design of antireflection surfaces and thin materials with efficient light absorption. Since for practical realization of a such medium, one need to consider the effect of truncation of the profile (finite spatial extent), here we have addressed this issue. The results of this study yields new insights on wave propagation in this new class of optical media, and may also provide guidance in the development of metamaterials based on appropriately chosen permittivity profile. The on-demand delay-controlled suppression of scattering on one side of the structure and the unidirectional excitation of the surface plasmons may also find applications for optical logic operations. With the rapid development in the field of dispersion engineering, metamaterials research and nano-fabrication technology, it is hoped that fabrication and practical realization of this new class of optical media will be possible in the near future. This would enable observation of a number of fundamental effects as predicted here and in other contemporary research, and may eventually lead to the development of novel optical devices. This theoretical study may therefore stimulate experimental research on this emerging new class of optical materials.
\section*{Acknowledgements}
Authors would like to thank Shourya Dutta-Gupta for help in preparing the manuscript and for many helpful discussions. SS thanks Indian Institute of Science Education and Research Kolkata (IISER-K) for the research fellowship and the School of Physics, University of Hyderabad (UoH) for providing local hospitality during his stay at the UoH campus.

\section*{Referencs}

\bibliographystyle{iopart-num}
\bibliography{cparef}

\providecommand{\newblock}{}
\begin{thebibliography}{10}
\expandafter\ifx\csname url\endcsname\relax
  \def\url#1{{\tt #1}}\fi
\expandafter\ifx\csname urlprefix\endcsname\relax\def\urlprefix{URL }\fi
\providecommand{\eprint}[2][]{\url{#2}}

\bibitem{bornwolf}
Born M and Wolf E 1999 {\em Principles of Optics\/} 7th ed (Cambridge
  University Press)

\bibitem{Leknerbook}
Lekner J 1987 {\em Theory of Reflection of Electromagnetic and Particle
  Waves\/} (Springer)

\bibitem{poitras2004}
Poitras D and Dobrowolski J 2004 {\em Applied optics\/} {\bf 43} 1286--1295

\bibitem{Kiriuscheva}
Kiriuscheva N and Kuzmin S 1998 {\em Am. J. Phys.\/} {\bf 66} 867--872

\bibitem{lekner2007reflectionless}
Lekner J 2007 {\em Am. J. Phys\/} {\bf 75} 1151--1157

\bibitem{Baryshnikova2015}
Baryshnikova K~V, Kadochkin A~S and Shalin A~S 2015 {\em Optics and
  Spectroscopy\/} {\bf 119} 343--355 ISSN 1562-6911

\bibitem{Horsleywave2016}
Horsley S~A~R, King C~G and Philbin T~G 2016 {\em Journal of Optics\/} {\bf 18}
  044016 ISSN 2040-8986

\bibitem{HorsleyKdV2016}
Horsley S~A~R 2016 {\em Journal of Optics\/} {\bf 18} 085104 ISSN 2040-8986

\bibitem{kay1956}
Kay I and Moses H~E 1956 {\em J. Appl. Phys.\/} {\bf 27} 1503--1508

\bibitem{sdgoe2007}
Gupta S~D and Agarwal G~S 2007 {\em Opt. Express\/} {\bf 15} 9614--9624

\bibitem{sdgoe2014}
Thekkekara L~V, Achanta V~G and Gupta S~D 2014 {\em Opt. Express\/}

\bibitem{Horsleynatph2015}
Horsley S~A~R, Artoni M and La~Rocca G~C 2015 {\em Nat Photon\/} {\bf 9}
  436--439 ISSN 1749-4885

\bibitem{LonghiOL2015}
Longhi S 2015 {\em Opt. Lett.\/} {\bf 40} 5694--5697

\bibitem{LonghiEPL2015}
Longhi S 2015 {\em EPL (Europhysics Letters)\/} {\bf 112} 64001 ISSN
  `0295-5075'

\bibitem{Mostafa2014}
Mostafazadeh A 2014 {\em Phys. Rev. A\/} {\bf 89} 012709

\bibitem{waveopticsbook}
Gupta S~D, Ghosh N and Banerjee A 2015 {\em Wave optics: Basic Concepts and
  Contemporart Trends\/} (CRC Press)

\bibitem{Longhi2011}
Longhi S 2011 {\em J. Phys. A: Math. Theor.\/} {\bf 44} 485302

\bibitem{Feng2013}
Feng L, Xu Y~L, Fegadolli W~S, Lu M~H, Oliveira J~E~B, Almeida V~R, Chen Y~F
  and Scherer A 2013 {\em Nat Mater\/} {\bf 12} 108--113 ISSN 1476-1122

\bibitem{prlLin2011}
Lin Z, Ramezani H, Eichelkraut T, Kottos T, Cao H and Christodoulides D~N 2011
  {\em Phys. Rev. Lett.\/} {\bf 106} 213901

\bibitem{Longhi_bidirection}
Longhi S 2016 {\em Opt. Lett.\/} {\bf 41} 3727--3730

\bibitem{wan2011}
Wan W, Chong Y~D, Ge L, Noh H, Stone A~D and Cao H 2011 {\em Science\/} {\bf
  331} 889�892

\bibitem{sdgOE2012}
Dutta-Gupta S, Martin O~J~F, Gupta S~D and Agarwal G~S 2012 {\em Opt.
  Express\/} {\bf 20} 1330--1336

\bibitem{sdgOL2012}
Dutta-Gupta S, Deshmukh R, Gopal A~V, Martin O~J~F and Gupta S~D 2012 {\em Opt.
  Lett.\/} {\bf 37} 4452--4454

\bibitem{sdgOL2007}
Gupta S~D 2007 {\em Opt. Lett.\/} {\bf 32} 1483--1485

\bibitem{nireekshan_absorption2013}
Reddy K~N and Gupta S~D 2013 {\em Opt. Lett.\/} {\bf 38} 5252--5255

\bibitem{reddyCC2013}
Reddy K~N, Gopal A~V and Gupta S~D 2013 {\em Opt. Lett.\/} {\bf 38} 2517--2520

\bibitem{gapsoliton2014}
Reddy K~N and Gupta S~D 2014 {\em Opt. Lett.\/} {\bf 39} 2254--2257

\bibitem{twophotonSDG}
Gupta S~D and Agarwal G~S 2014 {\em Opt. Lett.\/} {\bf 39} 390--393

\bibitem{gsa_cpa}
Huang S and Agarwal G~S 2014 {\em Opt. Express\/} {\bf 22} 20936--20947

\bibitem{Roger2015}
Roger T, Vezzoli S, Bolduc E, Valente J, Heitz J~J~F, Jeffers J, Soci C, Leach
  J, Couteau C, Zheludev N~I and Faccio D 2015 {\em Nat. Commun.\/} {\bf 6}
  7031

\bibitem{receprocity_sdg2002}
Agarwal G~S and Gupta S~D 2002 {\em Opt. Lett.\/} {\bf 27} 1205--1207

\bibitem{sdg_joptb2004}
Rao V~S~C~M, Gupta S~D and Agarwal G~S 2004 {\em J. Opt. B: Quantum Semiclass.
  Opt.\/} {\bf 6} 555�562

\bibitem{Jalas_2013}
Jalas D, Petrov A, Eich M, Freude W, Fan S, Yu Z, Baets R, Popovi{\'{c}} M,
  Melloni A, Joannopoulos J~D, Vanwolleghem M, Doerr C~R and Renner H 2013 {\em
  Nat Photon\/} {\bf 7} 579--582

\bibitem{Potton_2004}
Potton R~J 2004 {\em Rep. Prog. Phys.\/} {\bf 67} 717--754

\bibitem{PTsymm_sdg}
Liu X, Gupta S~D and Agarwal G~S 2014 {\em Phys. Rev. A\/} {\bf 89} 013824

\bibitem{epjap_2016}
Mukherjee S and Gupta S~D 2016 {\em The European Physical Journal Applied
  Physics\/} {\bf 76} 30001

\bibitem{Kumari_2012}
Kumari M and Gupta S~D 2012 {\em Opt. Commun.\/} {\bf 285} 617--620

\bibitem{Woerdman_2003}
Altewischer E, van Exter M~P and Woerdman J~P 2003 {\em Opt. Lett.\/} {\bf 28}
  1906

\end{thebibliography}





\end{document}